\documentclass[preprintnumbers,floatfix,aps,prd,preprint]{revtex4}

\usepackage{amsmath}
\usepackage{amsfonts}
\usepackage{amssymb}
\usepackage{bm}
\usepackage{enumerate}
\usepackage[dvips]{color,graphicx}
\usepackage{epsfig}
\usepackage[figuresright]{rotating}

\newcommand{\eps}{\varepsilon}

\preprint{JLAB-THY-14-1843}

\begin{document}

\title{Quasielastic electron--deuteron scattering in the \\
	weak binding approximation}

\author{J.~J.~Ethier$^{1,2}$,
	N.~Doshi$^{1,3}$,
	S.~Malace$^1$,
	W.~Melnitchouk$^1$}
\affiliation{
$^1$\mbox{Jefferson Lab, 12000 Jefferson Avenue,
	  Newport News, Virginia 23606, USA}		\\
$^2$\mbox{Department of Physics, College of William and Mary,
	  Williamsburg, Virginia 23185, USA}		\\
$^3$\mbox{Department of Physics, Carnegie Mellon University,
	  Pittsburgh, Pennsylvania 15213, USA}}

\begin{abstract}
We perform a global analysis of all available electron--deuteron
quasielastic scattering data using $Q^2$-dependent smearing functions
that describe inclusive inelastic $ed$ scattering within the weak
binding approximation.  We study the dependence of the cross sections
on the deuteron wave function and the off-shell extrapolation of the
elastic electron--nucleon cross sections, which show particular
sensitivity at $x \gg 1$.
The excellent overall agreement with data over a large range of $Q^2$
and $x$ suggests a limited need for effects beyond the impulse
approximation, with the possible exception of the very high-$x$
or very low-$Q^2$ regions, where short-distance effects or scattering
from non-nucleonic constituents in the deuteron become more relevant.
\end{abstract}

\date{\today}
\maketitle

\section{Introduction}
\label{sec:intro}

The deuteron has long been recognized as an ideal laboratory for
studying the dynamics of nucleon--nucleon ($NN$) interactions.
In particular, when the four-momentum transfer squared $Q^2$ is of
the order of the nucleon mass squared, $M^2$, or when the fraction
of momentum $x$ carried by the scattered quarks in the deuteron
is in the vicinity of $x \sim 1$, the role of short-distance
effects in the deuteron wave function becomes more prominent.
This region makes it possible to explore the structure of the
simplest nuclear bound state directly from the underlying theory
of the strong interactions, QCD.
Together with the constraints on the long-range structure of the
deuteron derived from chiral effective theory, the ultimate goal,
of course, is to obtain a quantitative description of the deuteron's
structure over all distance scales.

From a more practical perspective, experiments involving electron
(or other lepton) scattering from the deuteron targets have provided
the main source of information about the structure of the neutron.
The absence of free neutron targets has meant that properties such
as the neutron's elastic form factors or deep-inelastic structure
functions are usually extracted from measurements involving deuterons,
using empirical information about the corresponding proton
observables and knowledge of the $NN$ interaction in the deuteron.
Use of heavier nuclei necessarily increases the size of the bound
state effects, exacerbating the uncertainties introduced through
our incomplete knowledge of the nuclear wave function and the
reaction mechanism.

A robust extraction of neutron information requires
a reliable baseline model which accounts for the standard nuclear
physics in the deuteron.  This is usually embodied in the nuclear
impulse approximation, in which the probe scatters incoherently from
individual nucleons in the deuteron \cite{West71, Jaffe85, Jaffe89}.
Corrections to this framework arise in the form of rescattering
or final state interactions between the struck nucleon and the
spectator recoil \cite{Cosyn13, Cosyn11, Ciofi01, Ciofi04, Ciofi09},
as well as meson exchange currents \cite{Kaptari91, MT93}, nucleon
off-shell corrections \cite{GL92, MST94, MSTPLB}, and possible
non-nucleonic components of the deuteron wave function.
The unambiguous identification of these more exotic effects is
only feasible when the baseline calculations within the impulse
approximation can be performed with a sufficient degree of precision.

A successful framework which has been used to describe inclusive
inelastic electron scattering from nuclei is the weak binding
approximation (WBA), developed by \mbox{Kulagin {\it et~al.}}
and applied to both unpolarized \cite{KPW94, AKL04, KP06, Kahn09}
and polarized \cite{KMPW95, KMd, Kahn09} scattering from the
deuteron, as well as $^3$He \cite{KM3He, EM13} and heavier
nuclei \cite{KP06}.
It was also utilized in the extraction of the free neutron structure
function $F_2^n$ from inclusive deuterium and proton data in the
nucleon resonance region \cite{Malace09}, and the subsequent
verification of quark-hadron duality in the neutron
\cite{Malace10, MEK05}.

Of course, any general approach which aspires to have predictive
power must be able to describe a wider set of observables than
just a limited class of reactions.  Perhaps the most direct window
on the nuclear structure of the deuteron is offered by the process
of quasi-elastic (QE) scattering, where the electron scatters
elastically from a proton or neutron bound in the deuteron.
A large body of data has been accumulated on QE electron--deuteron
scattering over the past few decades, covering a large range of $Q^2$
(between $Q^2 \approx 0.1$ and 10~GeV$^2$) and energies $E$
(between $E \approx 0.2$ and 20~GeV), at forward and backward
scattering angles \cite{Database, Benhar08}.

In the impulse approximation the QE cross section is proportional to
the light-cone momentum distributions of nucleons in the deuteron,
$f^{N/d}$ (also referred to as the ``smearing functions'').
These are the same distributions that are used to compute the deuteron
structure functions in deep-inelastic scattering \cite{West71, Jaffe85,
MST94, MSTPLB, KPW94, KP06}, where they are convoluted with the
inelastic structure functions of the bound nucleons.  The resulting
convolutions depend rather sensitively on the precise structure
of both the smearing functions and the nucleon structure functions.
For QE scattering the deuteron structure functions are directly
given by $f^{N/d}$, multiplied by $Q^2$-dependent elastic nucleon
form factors.  This makes QE scattering the ideal testing ground
for models of the deuteron structure and the details of the nucleon
momentum distributions.

Despite the extensive work that has been carried out on computing
the smearing functions for application to deep-inelastic scattering,
using realistic deuteron wave functions and including finite-$Q^2$
corrections, surprisingly there has never been a direct test of
the WBA formalism with QE scattering data.  In this paper we
perform such an analysis, confronting the calculated light-cone
momentum distributions with the entire set of available QE cross
sections.  The level of agreement between the data and theory will
reveal the limits of validity of the WBA in the impulse approximation,
and the degree to which rescattering or more exotic effects need to
be incorporated for a complete description of electron--deuteron
scattering.

In Sec.~\ref{sec:ia} we review the formalism needed to describe
electron--deuteron scattering in the QE region, and outline the
derivation of the unpolarized deuteron $F_1^d$ and $F_2^d$ structure
functions within the WBA.  We examine the possible effects of the
modification of nucleon structure functions off the mass-shell, and
estimate the uncertainty on this modification using two different
prescriptions for the electromagnetic current commonly invoked in
the literature.  The calculated QE cross sections are compared
in Sec.~\ref{sec:results} with all available data on inclusive
electron--deuteron scattering in the QE region, for $x \gtrsim 1$.
We compare the predictions using the same smearing functions
as those utilized in deep-inelastic scattering, including the
kinematical, finite-$Q^2$ corrections to the smearing functions
derived in the high-$Q^2$ limit.
We further investigate the dependence of the cross sections
on the deuteron wave function, for several models based on
high-precision $NN$ potentials, as well as on the effects
of the nucleon off-shell corrections.
A comprehensive comparison with the data such as this will allow
us to clearly delineate the regions where the impulse approximation
is adequate for understanding the essential features of the data,
and identify where additional effects may be needed in future
analyses ($x \gg 1$).
Finally, in Sec.~\ref{sec:conc} we summarize our findings and
discuss their implications for future work.

\section{Quasi-Elastic Scattering in the Impulse Approximation}
\label{sec:ia}

In this section we summarize the main results for inclusive
electron--deuteron scattering in the impulse approximation.
After reviewing the general results for the deuteron structure
functions within the framework of the WBA, we describe how the
results are applied to the case of elastic scattering from
the nucleon bound inside the deuteron.
We present results for the case where the bound nucleon structure
is assumed to be the same as that for a free nucleon, as well as
for the more general case where the off-shell structure of the
bound nucleon is explicitly taken into account.

\subsection{Inclusive cross section and structure functions}

The inclusive cross section for the scattering of an incident electron
(with four-momentum $k_\mu$) from a deuteron target ($P_\mu$) to a
recoil electron ($k'_\mu$) and unobserved hadronic state $X$,
$e d \to e X$, is given in the target rest frame by
\begin{eqnarray}
{ d^2\sigma \over d\Omega dE' }
&=& {\alpha^2 \over Q^4} {E' \over E} {1 \over M_d}
    L_{\mu\nu}\, W^{\mu\nu},
\label{eq:sigLW}
\end{eqnarray}
where $\alpha$ is the electromagnetic fine structure constant,
$E\, (E')$ is the incident (scattered) electron energy, and
$M_d$ is the deuteron mass.
The invariant mass squared of the exchanged photon is given by
$Q^2 \equiv -q^2 \approx 4 E E' \sin^2(\theta/2)$, where $\theta$
is the electron scattering angle in the target rest frame, with
$q_\mu = k_\mu - k'_\mu$ the exchanged photon four-momentum.
The leptonic tensor in Eq.~(\ref{eq:sigLW}) is given by
\begin{eqnarray}
L_{\mu\nu}
&=& 2 k_\mu k'_\nu + 2 k'_\mu k_\nu + q^2 g_{\mu\nu},
\end{eqnarray}
while the deuteron hadronic tensor $W^{\mu\nu}$ is parametrized
by the deuteron structure functions $F_1^d$ and $F_2^d$,
\begin{eqnarray}
W^{\mu\nu}(P,q)
&=& \left( -g^{\mu\nu} + {q^\mu q^\nu \over q^2} \right)
     F_1^d\
 +\ \left( P^\mu - {P \cdot q \over q^2} q^\mu \right)
    \left( P^\nu - {P \cdot q \over q^2} q^\nu \right)
    {F_2^d \over P \cdot q}.
\label{eq:Wmunu_gen}
\end{eqnarray}
In terms of the deuteron structure functions, which are usually
expressed as functions of $Q^2$ and the Bjorken scaling variable
$x=Q^2/2M\nu$, where $\nu=E-E'$ is the energy transfer, the
inclusive cross section can then be written as
\begin{eqnarray}
\label{eq:sigma}
{ d^2\sigma \over d\Omega dE' }
&=& \sigma_{\rm Mott}
    \left( {2\over M_d} \tan^2{\theta\over 2} F_1^d(x,Q^2)
	 + {1\over\nu} F_2^d(x,Q^2)
    \right),
\end{eqnarray}
where $\sigma_{\rm Mott} = (4\alpha^2 E'^2/Q^4) \cos^2(\theta/2)$
is the Mott cross section for scattering from a point particle.
Note that at forward scattering angles ($\theta \to 0^\circ$) the
cross section is given entirely by the $F_2^d$ structure function,
while at backward angles ($\theta \to 180^\circ$) the $F_1^d$
structure function is dominant.

\subsection{Weak binding approximation}

To relate the deuteron cross section or structure functions to
those of the nucleon requires modeling of the distribution and
interaction of the bound nucleons in the deuterium nucleus.
Within a covariant framework the deuteron hadronic tensor $W^{\mu\nu}$
in Eq.~(\ref{eq:Wmunu_gen}) can be written as a product of the
nucleon--deuteron scattering amplitude $\widehat{{\cal A}}$ and the
truncated nucleon hadronic tensor $\widehat{{\cal W}}_N^{\, \mu\nu}$
describing the structure of the off-shell nucleon \cite{MST94},
\begin{eqnarray}
W^{\mu\nu}(P,q)
&=& \int\!\!{ d^4p \over (2\pi)^4 }
    {\rm Tr}\left[ \widehat{{\cal A}}(P,p)\,
		   \widehat{{\cal W}}_N^{\, \mu\nu}(p,q)
	    \right],
\end{eqnarray}
where $p$ is the four-momentum of the struck nucleon.
Expanding the nuclear amplitude $\widehat{{\cal A}}$ to order
$\bm{p}^2/M^2$ in the bound nucleon three-momentum and to order
$\eps/M$ in the energy $\eps \equiv p_0 - M$, the deuteron
tensor simplifies to an integral over the nonrelativistic
deuteron spectral function ${\cal P}$ and the nucleon hadronic
tensor $W_N^{\mu\nu}$ \cite{KP06},
\begin{eqnarray}
W^{\mu\nu}(P,q)
&=& \int\!\!{ d^4p \over (2\pi)^4 } {M_d \over M+\eps}\
    {\cal P}(\eps,\bm{p})\, W_N^{\mu\nu}(p,q)\
 +\ {\cal O}(|\bm{p}|^3/M^3).
\label{eq:Wmunu}
\end{eqnarray}
The spectral function is written in terms of the deuteron wave
function $\psi_d$ as
\begin{eqnarray}
{\cal P}(\eps,\bm{p})
&=& (4\pi^3)\,
  \delta\left(\eps - \eps_d + {\bm{p}^2 \over 2M}\right)\,
  \left| \psi_d({\bm p}) \right|^2,
\label{eq:specfn}
\end{eqnarray}
where the deuteron binding energy $\eps_d = M_d-2M$
and the wave function is normalized according to 
$\int d^3{\bm p}\, |\psi_d({\bm p})|^2 = 4\pi$.

Evaluating explicitly the hadronic tensor in Eq.~(\ref{eq:Wmunu})
with the spectral function in Eq.~(\ref{eq:specfn}), and equating
the coefficients of the tensor in Eq.~(\ref{eq:Wmunu_gen}), one can
write the deuteron $F_1^d$ and $F_2^d$ structure functions in the
WBA in terms of the deuteron wave function $\psi_d({\bm p})$ and the
bound nucleon structure functions $\widetilde{F}_1^N$
and $\widetilde{F}_2^N$ \cite{AKL04, KP06, Kahn09, AVQ},
\begin{subequations}
\label{eq:F12QEful}
\begin{eqnarray}
x F_1^d(x,Q^2)
&=& \sum_N
    \int\!{d^3{\bm p} \over (2\pi)^3}
    \left| \psi_d({\bm p}) \right|^2
    \left( 1 + \frac{\gamma p_z}{M} \right)
    \left[ {\cal C}_{11}\,
	   {x \over y} \widetilde{F}_1^N\left({x \over y},Q^2,p^2\right)
	 + {\cal C}_{12}\,
	   \widetilde{F}_2^N\left({x \over y},Q^2,p^2\right)
   \right],
\label{eq:F1QEful}				\nonumber\\
& &						\\
F_2^d(x,Q^2)
&=& \sum_N
    \int\!{d^3{\bm p} \over (2\pi)^3}
    \left| \psi_d({\bm p}) \right|^2
    \left( 1 + \frac{\gamma p_z}{M} \right)
    {\cal C}_{22}\, \widetilde{F}_2^N\left({x \over y},Q^2,p^2\right),
\label{eq:F2QEful}
\end{eqnarray}
\end{subequations}
where $\gamma^2 = 1 + 4 M^2 x^2/Q^2$ is a kinematical factor,
and the sum runs over $N=p$ and $n$.
The variable $y$ is the light-cone momentum fraction of the
deuteron carried by the interacting nucleon,
\begin{eqnarray}
y &=& {M_d \over M} {p\cdot q \over P\cdot q}\
   =\ {p_0 + \gamma p_z \over M},
\label{eq:y}
\end{eqnarray}
and the coefficients ${\cal C}_{ij}$ are given by
\begin{subequations}
\label{eq:Cij}
\begin{eqnarray}
{\cal C}_{11}
&=& 1,						\\
{\cal C}_{12}
&=& (\gamma^2-1) {{\bm p}_\perp^2 \over 4 y^2 M^2},	\\
{\cal C}_{22}
&=& {1 \over \gamma^2}
\left[
1 + {(\gamma^2-1) \over 2 y^2 M^2}
    \left( 2 p^2 + 3 {\bm p}_\perp^2 \right)
\right].
\end{eqnarray}
\end{subequations}
Because the struck nucleon is off its mass shell with virtuality
$p^2 = p_0^2 - \bm{p}^2 < M^2$, where the interacting nucleon's
energy is $p_0 = M_d - \sqrt{M^2 + \bm{p}^2}$, the structure functions
$\widetilde{F}_1^N$ and $\widetilde{F}_2^N$ in Eqs.~(\ref{eq:F12QEful})
can in principle also depend on $p^2$, in addition to $x$ and $Q^2$.
In practice, since the binding energy is a small ($\approx 0.1\%$)
fraction of the deuteron's mass, and the average nucleon momentum
in the deuteron is $|\bm{p}| \sim 130$~MeV, the typical nucleon
virtuality $(p^2)^{1/2}$ is only $\sim 2\%$ less than the free
nucleon mass.  As a reasonable first approximation, therefore,
one can take the bound nucleon structure functions to be the same
as their on-shell limits,
$\widetilde{F}_{1,2}^N(x,Q^2,p^2)
 \approx \widetilde{F}_{1,2}^N(x,Q^2,M^2)
 \equiv F_{1,2}^N(x,Q^2)$.
In this case the $p^2$ (or ${\bm p}_\perp^2$) and $y$ dependence in
Eqs.~(\ref{eq:F12QEful}) factorizes, and the integration can be
reduced to a one-dimensional convolution in $y$ \cite{Kahn09},
\begin{subequations}
\label{eq:conv}
\begin{eqnarray}
xF_1^d(x,Q^2)
&=& \sum_N \int_x^{M_d/M}\!dy
    \left[ f_{11}^{N/d}(y,\gamma)\,
	   {x \over y} F_1^N\left({x \over y},Q^2\right)
	 + f_{12}^{N/d}(y,\gamma)\,
	   F_2^N\left({x \over y},Q^2\right)
    \right],\ \ \ \
\label{eq:conv1}				\\
F_2^d(x,Q^2)
&=& \sum_N \int_x^{M_d/M}\!dy
    \left[ f_{22}^{N/d}(y,\gamma)\,
	   F_2^N\left({x \over y},Q^2\right)
    \right],
\label{eq:conv2}
\end{eqnarray}
\end{subequations}
where the nucleon smearing functions in the deuteron
$f_{ij}^{p/d} = f_{ij}^{n/d} \equiv f_{ij}$ (assuming isospin symmetry)
are given by \cite{AKL04, KP06, Kahn09, AVQ}
\begin{eqnarray}
f_{ij}(y,\gamma)
&=& \int\!{d^3{\bm p} \over (2\pi)^3}
    \left| \psi_d({\bm p}) \right|^2
    \left( 1 + \frac{\gamma p_z}{M} \right)
    {\cal C}_{ij}\,
    \delta\left( y-1-\frac{\eps+\gamma p_z}{M} \right).
\label{eq:fij}
\end{eqnarray}
In the $\gamma \to 1$ limit the functions $f_{ij}$ can be
interpreted as light-cone momentum distributions of nucleons
in the deuteron, giving the probability of finding a nucleon
with a light-cone momentum fraction $y$ inside the deuteron.
For $\gamma=1$ the smearing functions are therefore normalized as
\begin{eqnarray}
\label{eq:norm}
\int_0^{M_d/M} dy\, f_{ii}(y,1) &=& 1, \hspace*{1cm}
\int_0^{M_d/M} dy\, f_{12}(y,1)\ =\ 0.
\end{eqnarray}
In this limit the convolutions for $F_1^d$ and $F_2^d$ are thus
diagonal in the structure function type, since ${\cal C}_{12} \to 0$
as $\gamma \to 1$.  At finite values of $Q^2$ the normalizations
(\ref{eq:norm}) no longer hold, and the distributions do not have
a probabilistic interpretation.
However, in practical calculations it is nonetheless vital
to keep the full $Q^2$ dependence of the smearing functions.

\begin{figure}
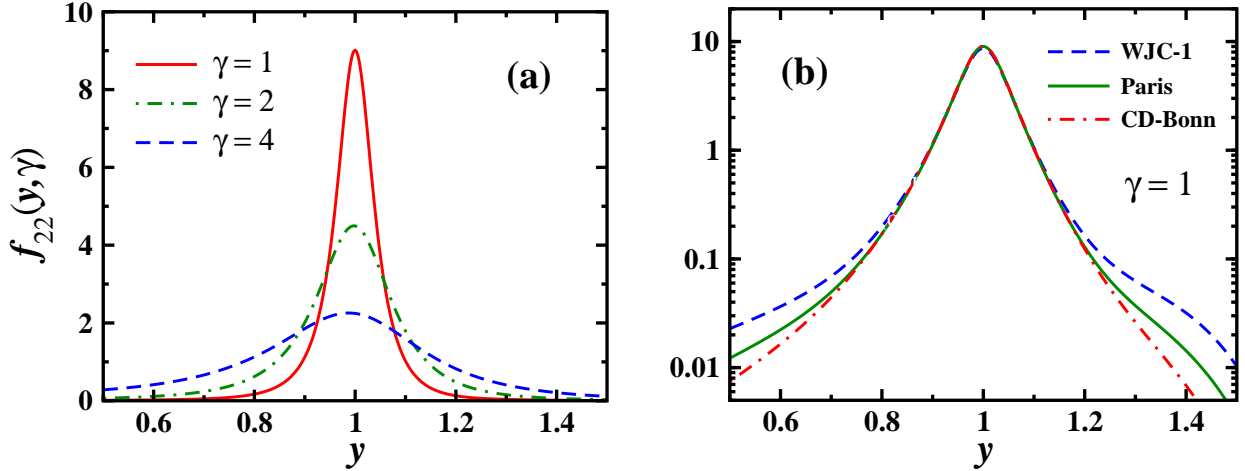

\includegraphics[width=8.0cm]{fig_fy.eps}\ \ \ \ \ \
\includegraphics[width=7.6cm]{fig_fy_log.eps}
\caption{Nucleon smearing function in the deuteron $f_{22}(y,\gamma)$
	as a function of $y$ for
	{\bf (a)} different values of $\gamma$ ($\gamma=1, 2, 4$)
	using the Paris \cite{Paris} wave function, and
	{\bf (b)} $\gamma=1$ using the Paris, WJC-1 \cite{WJC}
	and CD-Bonn \cite{CDBonn} wave functions.}
\label{fig:fy}
\end{figure}

In Fig.~\ref{fig:fy} the smearing function $f_{22}(y,\gamma)$ relevant
for the $F_2^d$ structure function is illustrated for different values
of $\gamma$ and for different models of the deuteron wave function.
In the $Q^2 \to \infty$ limit, the function is strongly peaked
around $y=1$, with a maximum value of $\approx 9$, but becomes
broader with increasing $\gamma$, with the peak about half as large
for $\gamma=2$ and 1/4 as large for $\gamma=4$ compared with that
in the scaling limit.
Note that at $x=1$ the value of $\gamma$ is $\approx 4.3$, 2.1 and 1.2
at $Q^2 = 0.2$, 1 and 10~GeV$^2$, respectively, which covers most of
the $Q^2$ range of the available QE data.
The behavior of the $f_{11}(y,\gamma)$ smearing function is
qualitatively similar to that in Fig.~\ref{fig:fy}.

At $y \approx 1$ the smearing function is determined mostly by
the long-distance part of the deuteron wave function, which has
relatively weak model dependence.  The tails of the distributions
at $|y-1| \gg 0$, on the other hand, exhibit strong deuteron model
dependence, with the WJC-1 wave function \cite{WJC} giving the
hardest distribution (largest tails in $f_{22}$), the CD-Bonn
\cite{CDBonn} the softest distribution (smallest $f_{22}$),
and the Paris wave function intermediate between the two.
These features will be directly reflected in the model dependence
of the structure function contributions to the QE cross sections
in Sec.~\ref{sec:results}.

Note that whereas some earlier analyses of QE and inelastic
electron--deuteron scattering made use of {\it ad hoc}
assumptions and {\it ans\"{a}tze} (see Ref.~\cite{MST94} for
a discussion), the expressions in Eqs.~(\ref{eq:F12QEful})
are systematically expanded in the bound nucleon momentum,
and are exact to order $\bm{p}^2/M^2$ for all values of $Q^2$.
As discussed above, the only assumption that has been made in
Eqs.~(\ref{eq:conv}) is that the bound nucleon structure functions
appearing in the convolutions are not modified off-shell.
The latter assumption constitutes one of the largest sources
of theoretical uncertainty in the calculation of the deuteron
structure functions.
In Sec.~\ref{ssec:QEoff} we will explore the possible effects
of the $p^2$ dependence of the bound nucleon structure function
on the QE cross section.  Before doing so, however, we first
consider the specific case of elastic scattering from the nucleon.

\subsection{Quasi-elastic structure functions}
\label{ssec:QEon}

For electron scattering from a free nucleon, the matrix element
of the electromagnetic current operator $J^\mu$ for an elastic
final state [$(p+q)^2 = M^2$] is parametrized in terms of the 
Dirac $F_{1N}$ and Pauli $F_{2N}$ form factors (not to be confused
with the inclusive structure functions $F_{1,2}^N(x,Q^2)$,
which are always functions of {\it two} variables),
\begin{eqnarray}
\langle N(p+q) | J^\mu | N(p) \rangle
&=& \bar{u}(p+q)
    \left[ \gamma^\mu\, F_{1N}(Q^2)\
	+\ i \sigma^{\mu\nu} q_\nu {F_{2N}(Q^2) \over 2M}
    \right]
    u(p),
\label{eq:J1}
\end{eqnarray}
with the form factors normalized such that
$F_{1p}(0) = 1$, $F_{1n}(0) = 0$ and $F_{2N}(0) = \mu_N$,
where $\mu_N$ is the nucleon anomalous magnetic moment.
Using the Gordon identity for on-shell states, one can eliminate
the $\sigma^{\mu\nu}$ term in Eq.~(\ref{eq:J1}) to express the
matrix element of the electromagnetic current equivalently as
\begin{eqnarray}
\langle N(p+q) | J^\mu | N(p) \rangle
&=& \bar{u}(p+q)
    \left[ \gamma^\mu\, G_{MN}(Q^2)\
	-\ (2 p^\mu + q^\mu) {F_{2N}(Q^2) \over 2M}
    \right]
    u(p),
\label{eq:J2}
\end{eqnarray}
where $G_{MN}$ here is the Sachs magnetic form factor.
The Sachs electric and magnetic form factors are related
to the Dirac and Pauli form factors by
\begin{subequations}
\label{eq:ffs}
\begin{eqnarray}
F_{1N}(Q^2)
&=& {1 \over 1+\tau}
    \left[ G_{EN}(Q^2) + \tau G_{MN}(Q^2) \right],
\label{eq:ff1}					\\
F_{2N}(Q^2)
&=& {1 \over 1+\tau} 
    \left[ G_{MN}(Q^2) - G_{EN}(Q^2) \right].
\label{eq:ff2}
\end{eqnarray}
\end{subequations}
As we shall see in Sec.~\ref{ssec:QEoff} below, the expressions in
Eqs.~(\ref{eq:J1}) and (\ref{eq:J2}) are equivalent on-shell, but
can differ when the initial nucleon is off-shell.
In terms of the Sachs electric and magnetic form factors,
the elastic contributions to the inclusive structure functions
of a free nucleon are given by
\begin{subequations}
\label{eq:F12el}
\begin{eqnarray}
F_1^{N (\rm el)}(x,Q^2)
&=& \left[ {1 \over 2} G_{MN}^2(Q^2) \right]
    Q^2\ \delta\left((p+q)^2-M^2\right)
\label{eq:F1el}					\\
F_2^{N (\rm el)}(x,Q^2)
&=& \left[ {G_{EN}^2(Q^2) + \tau G_{MN}^2(Q^2) \over 1+\tau} \right]
    2 p\cdot q\ \delta\left((p+q)^2-M^2\right),
\label{eq:F2el}
\end{eqnarray}
\end{subequations}
where $\tau = Q^2/4M^2$.
Using the fact that for an on-shell nucleon ($p^2 = M^2$) one has
\mbox{$2 p \cdot q$} $= 4M^2 \tau$, and the $\delta$-functions in
Eqs.~(\ref{eq:F12el}) can also be written in terms of the $x$ variable,
$Q^2\, \delta\left((p+q)^2-M^2\right)
= 2 p\cdot q\, \delta\left((p+q)^2-M^2\right)
= \delta(1-x)$.
Substituting the elastic structure functions in Eqs.~(\ref{eq:conv}),
the deuteron QE structure functions can then be written as simple
products of the nucleon smearing functions $f_{ij}$ and the elastic
electromagnetic form factors,
\begin{subequations}
\label{eq:F12QE}
\begin{eqnarray}
x F_1^{d (\rm QE)}(x,Q^2)
&=& \sum_N
\left\{
    {1 \over 2} x f_{11}(x,\gamma)\,
    G_{M N}^2(Q^2)\
 +\ x f_{12}(x,\gamma)
    \left[ {G_{E N}^2(Q^2) + \tau G_{M N}^2(Q^2) \over 1+\tau}
    \right]
\right\},
\label{eq:F1QE}				\nonumber\\
& &					\\
F_2^{d (\rm QE)}(x,Q^2)
&=& \sum_N
    x f_{22}(x,\gamma)
    \left[ {G_{E N}^2(Q^2) + \tau G_{M N}^2(Q^2) \over 1+\tau}
    \right].
\label{eq:F2QE}
\end{eqnarray}
\end{subequations}
The $Q^2$ dependence of the QE structure functions arises from both
the $Q^2$ dependence of the elastic form factors and the $\gamma$
dependence of the smearing function.  The latter, as we shall see
in Sec.~\ref{ssec:comparison}, will in fact be vital for describing
the $Q^2$ dependence of QE cross section data.

\subsection{Nucleon off-shell corrections}
\label{ssec:QEoff}

For the case where the struck nucleon is bound inside the deuteron
and is thus off its mass-shell, $p^2 \neq M^2$, we can generalize the
elastic nucleon scattering contributions to the structure functions
by explicitly taking into account the kinematical $p^2$ dependence.
From the on-shell condition of the final nucleon, one has the constraint
$2 p\cdot q = Q^2 + M^2 - p^2 = Q^2/(x/y)$, where $y$ is defined in
Eq.~(\ref{eq:y}).  This enables the $\delta$-function in
Eqs.~(\ref{eq:F12el}) to be written as
	$ \delta\left((p+q)^2-M^2\right)
        = [(x/y) / Q^2]\, \delta(1-\kappa(p^2) x/y) $,
where $\kappa(p^2) = 1 + (M^2-p^2)/Q^2$.
Making use of the definition of the electromagnetic current in
Eq.~(\ref{eq:J1}), the elastic structure functions for the off-shell
nucleon are then given by
\begin{subequations}
\label{eq:off1}
\begin{eqnarray}
\widetilde{F}_1^{N (\rm el)}\Big({x\over y},Q^2,p^2\Big)
&=& \left[
      {G_{MN}^2 \over 2}
    - {(M^2-p^2) \over 2 Q^2}
      \left( {G_{EN}^2+\tau G_{MN}^2 \over 1+\tau}
	   - {(M^2-p^2) \over 4 M^2}
	     {(G_{MN}-G_{EN})^2 \over (1+\tau)^2}
      \right)
    \right]						\nonumber\\
& & \times\,
    {x \over y} \delta\Big(1-\kappa(p^2){x \over y}\Big),\\
\widetilde{F}_2^{N (\rm el)}\Big({x\over y},Q^2,p^2\Big)
&=& \left[ {G_{EN}^2 + \tau G_{MN}^2 \over 1+\tau}
    \right]
    \delta\Big(1-\kappa(p^2){x \over y}\Big).
\end{eqnarray}
\end{subequations}
This corresponds to what is known in the literature as the ``cc2''
prescription of De~Forest \cite{DeForest83}.

If one instead uses the form of the electromagnetic current in
Eq.~(\ref{eq:J2}), the elastic structure functions for the off-shell 
nucleon are given by the alternative forms
\begin{subequations}
\label{eq:off2}
\begin{eqnarray}
\widetilde{F}_1^{N (\rm el)}\Big({x\over y},Q^2,p^2\Big)
&=& \left[
      {G_{MN}^2 \over 2}
      \left( 1 - {M^2-p^2 \over Q^2} \right)
    \right]
    {x \over y} \delta\Big(1-\kappa(p^2){x \over y}\Big),\\
\widetilde{F}_2^{N (\rm el)}\Big({x\over y},Q^2,p^2\Big)
&=& \left[ {G_{EN}^2 + \tau G_{MN}^2 \over 1+\tau}
	 - {(M^2-p^2) \over 4M^2}
	   {(G_{MN}-G_{EN})^2 \over (1+\tau)^2}
    \right]
    \delta\Big(1-\kappa(p^2){x \over y}\Big).	\nonumber\\
& &
\end{eqnarray}
\end{subequations}
This form corresponds to the ``cc1'' prescription of
Ref.~\cite{DeForest83}.

While the on-shell limits of the two sets of expressions for the
structure functions in Eqs.~(\ref{eq:off1}) and (\ref{eq:off2})
are equivalent, off-shell these will give rise to numerically
different results for the QE cross sections.  These differences
will be an indication of the uncertainty in the calculation of
the deuteron cross section due to the off-shell extrapolation
of the nucleon hadronic tensor, which will be discussed in the
following section.

\section{Numerical results}
\label{sec:results}

Having derived the results for the contributions of the inclusive
deuteron structure functions to the QE cross section within the
framework of the WBA, we can now compare the predictions with
the available QE electron--deuteron scattering data.
In the following we first summarize the data sets used in this
analysis, before proceeding with the model comparisons.

\subsection{Electron--deuteron QE data sets}
\label{ssec:data}

QE electron--deuteron scattering cross sections have been measured
in a number of experiments at several facilities, including SLAC,
MIT-Bates, and Jefferson Lab, over a large range of energies and
scattering angles.  Most of these are summarized in the
``Quasielastic Electron Nucleus Scattering Archive'' \cite{Database},
which includes published data that have been radiatively corrected
and are not known to contain any pathologies.

The earliest data set was obtained by Schutz {\it et~al.} \cite{Schutz}
from SLAC in the late 1970s, containing forward scattering QE cross
sections at $\theta=8^\circ$ for incident energies between
$E \approx 6$~GeV and 18~GeV, and extending to very large values
of $x \lesssim 2$.
Backward angle data were obtained by Parker {\it et~al.}
\cite{Parker} at very low energies ($E \approx 0.2$~GeV)
from MIT-Bates, and by Arnold {\it et~al.} \cite{Arnold}
at higher energies ($E \approx 1$~GeV) from SLAC.
More extensive data sets from SLAC were collected in the early 1990s
by Lung \cite{Lung} around the QE peak for a range of scattering
angles $\theta \approx 15^\circ - 90^\circ$ at energies
$E \approx 1.5-5.5$~GeV, and by Rock  {\it et~al.} \cite{Rock} at
forward angles ($\theta=10^\circ$) at higher energies,
$E \approx 10-20$~GeV.  The latter offered access to the highest
available $Q^2$ values, reaching $Q^2=10$~GeV$^2$.
High precision data from Jefferson Lab were measured by Arrington
{\it et~al.} \cite{Arrington} at $\theta \approx 15^\circ - 50^\circ$
for energies between $E \approx 2$~GeV and 5~GeV, and most recently
by Fomin {\it et~al.} \cite{Fomin} in the vicinity of $x=1$ using
the 6~GeV CEBAF electron beam at angles between
$\theta \approx 18^\circ$ and $50^\circ$.

The complete QE data set amounts to over 2,000 data points covering
a range of $Q^2$ between $Q^2 \approx 0.1$ and 10~GeV$^2$ for energies
between $E \approx 0.2$ and 20~GeV, from $x \lesssim 1$ to $x \approx 2$.
In particular, the angular dependence of the cross sections allows the
effects of the $F_1^d$ and $F_2^d$ structure function contributions to
be studied independently.  Fitting these will constitute a significant
test of any model of the deuteron.

\subsection{Phenomenological analysis}
\label{ssec:comparison}

\begin{figure}
\includegraphics[width=17cm]{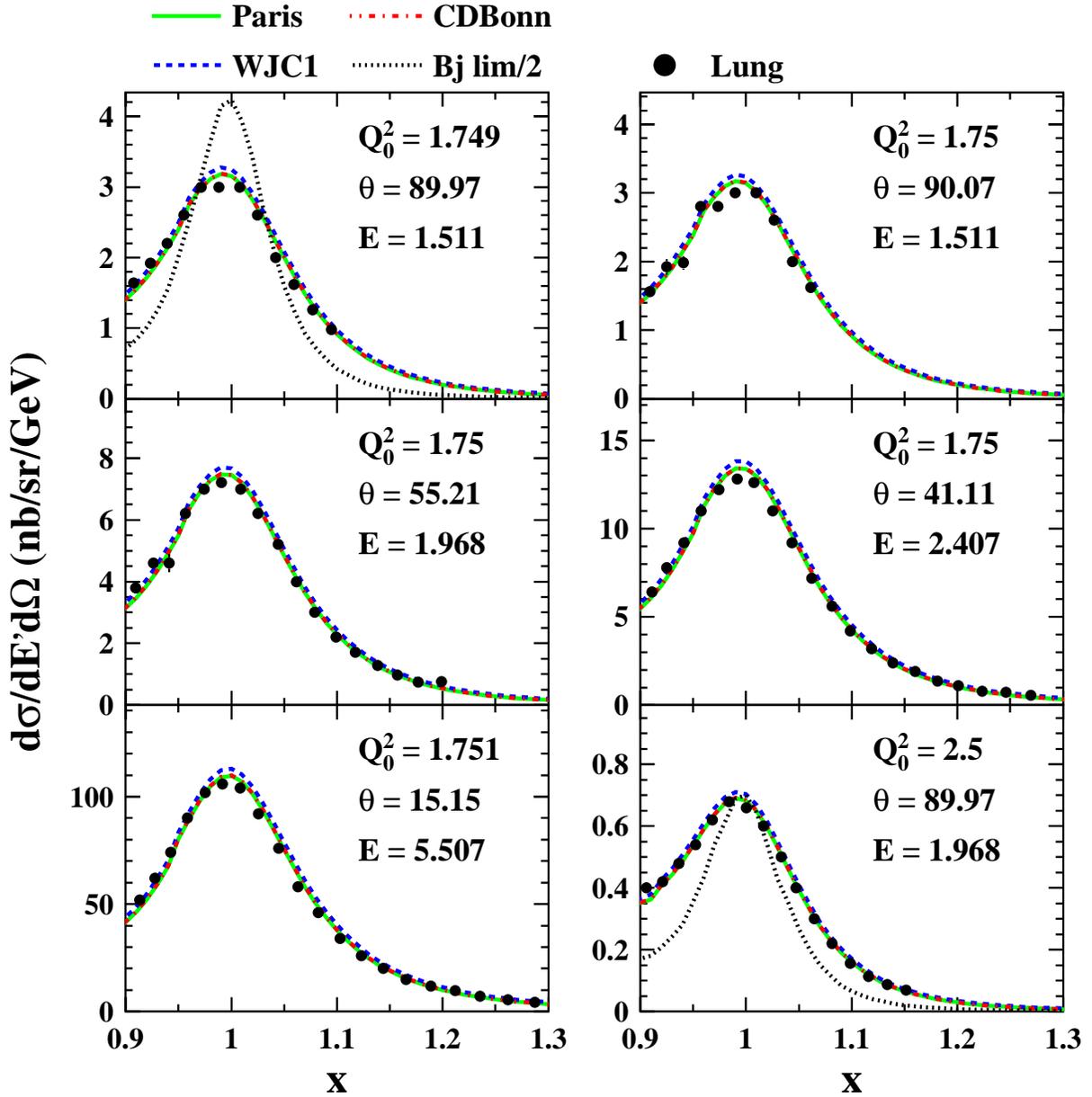}
\caption{Inclusive electron--deuteron scattering cross section in
	the QE region.  The SLAC data from Lung \cite{Lung}
	(filled circles) are compared with the WBA model predictions
	using the Paris \cite{Paris} (green solid curves), WJC-1
	\cite{WJC} (blue dashed curves) and CD-Bonn (red dot-dashed
        curves) \cite{CDBonn} deuteron wave functions.
	The results using the smearing functions computed in the
	large-$Q^2$ limit (black dotted curves) are also shown
	(scaled by a factor 1/2 for clarity).
	In this and subsequent figures, the energy $E$ (in GeV)
	and scattering angle $\theta$ (in degrees) are indicated
	on each panel; $Q_0^2$ (in GeV$^2$) is the value of the
	four-momentum transfer squared at $x=1$, which ranges
	here from $Q_0^2 \approx 1.75$ to 2.5~GeV$^2$.}
\label{fig:lung1}
\end{figure}

\begin{figure}
\includegraphics[width=17cm]{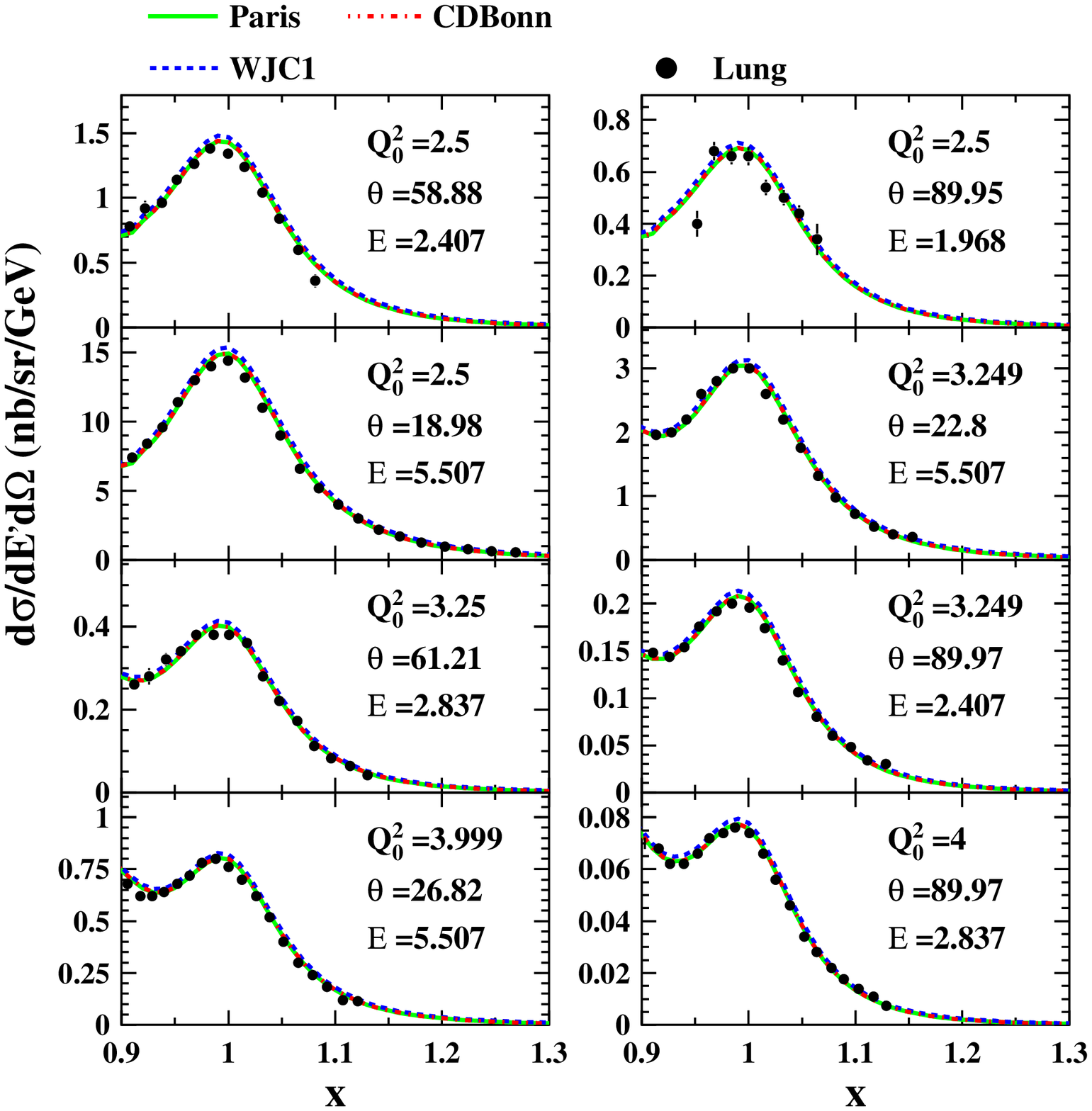}
\caption{As in Fig.~\ref{fig:lung1} but for $Q_0^2$ between
	2.5 and 4~GeV$^2$.}
\label{fig:lung2}
\end{figure}

Typical deuteron QE spectra are illustrated in Figs.~\ref{fig:lung1}
and \ref{fig:lung2}, where the cross sections are calculated in the
WBA model and compared with SLAC data from Lung \cite{Lung}.
The calculations were performed using several different deuteron
wave functions, based on the Paris \cite{Paris}, WJC-1 \cite{WJC}
and CD-Bonn \cite{CDBonn} nucleon--nucleon potentials,
and elastic nucleon form factors from the parametrizations
of Arrington {\it et~al.} \cite{AMT07} for the proton and
Bosted \cite{Bosted94} for the neutron.
Overall the agreement between the calculated cross sections
and the data is excellent.  This conclusion is independent of
the choice of input nucleon elastic form factors, with the results
using the parametrization of Kelly \cite{Kelly04} differing from
those in Figs.~\ref{fig:lung1} and \ref{fig:lung2} by $\lesssim 2\%$
for all kinematics.
Furthermore, in the $x$ range spanned by these data, $x \lesssim 1.2$,
the QE cross sections display very mild dependence on the deuteron
wave function.

In particular, the correct shape and magnitude of the QE peak is
well reproduced with the $y$- and $\gamma$-dependent smearing
functions of Eq.~(\ref{eq:fij}).
In contrast, using the smearing functions computed in the high-$Q^2$
($\gamma \to 1$) limit, as appropriate for deep-inelastic scattering
applications, the peak in the QE cross section would be a factor
of $\approx 2$ too large in the $Q^2$ range ($\sim 2$~GeV$^2$)
covered by the data in Figs.~\ref{fig:lung1} and \ref{fig:lung2}.
At significantly higher $Q^2$ ($\gtrsim 10$~GeV$^2$) the differences
between the full, finite-$Q^2$ results and the high-$Q^2$
approximation are reduced, but at values relevant to most of the
existing data the correct $Q^2$ dependence of the smearing functions
is vital to take into account.

\begin{figure}
\includegraphics[width=17cm]{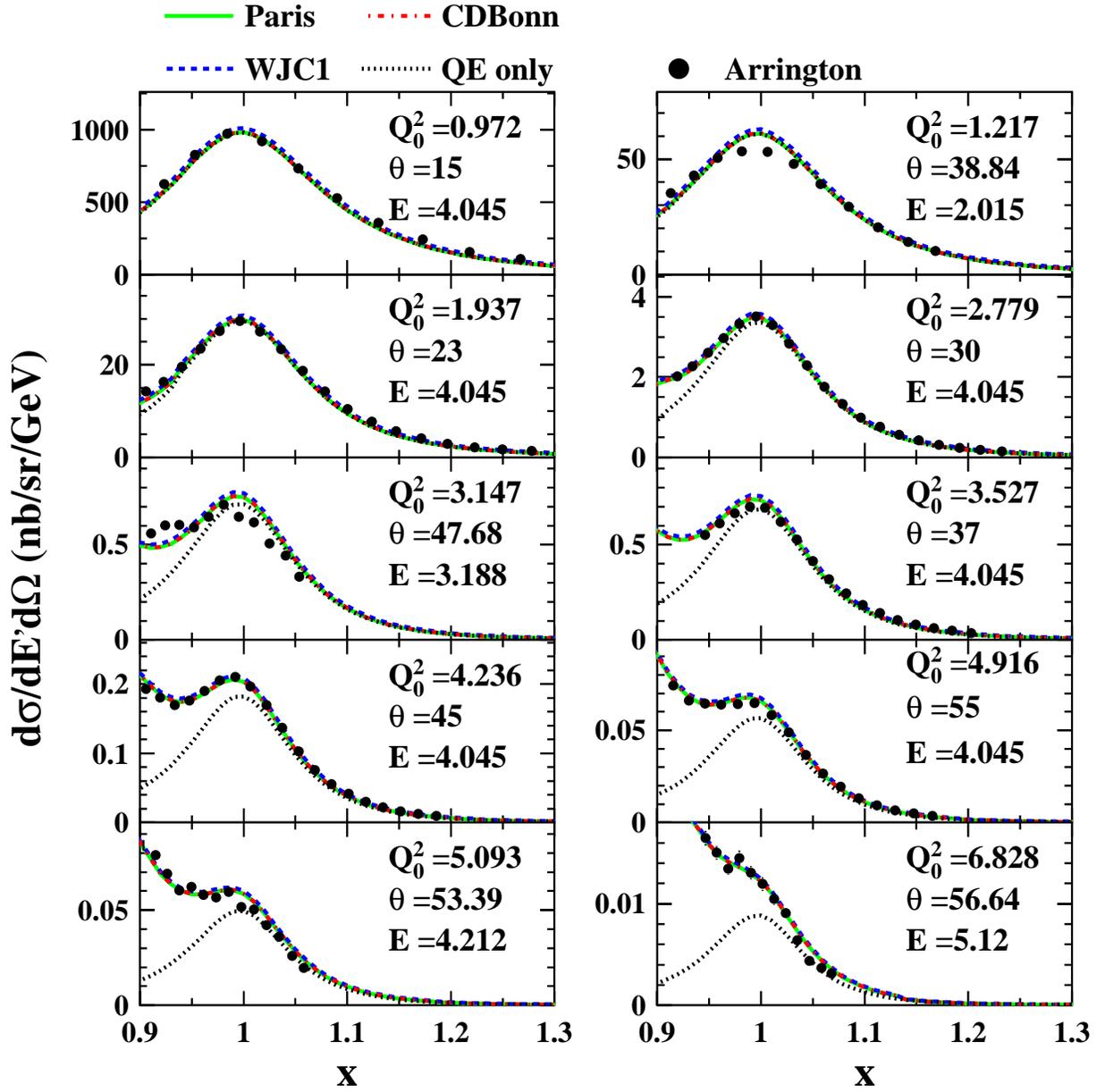}
\caption{Inclusive electron--deuteron QE scattering cross sections
	in the WBA model using the Paris \cite{Paris} (green solid
	curves), WJC-1 \cite{WJC} (blue dashed curves) and CD-Bonn
	(red dot-dashed curves) \cite{CDBonn} deuteron wave functions,
	compared with the Jefferson Lab data from Arrington
	{\it et~al.} \cite{Arrington}, for which $Q_0^2$ ranges
	between	$\approx 1$ and 7~GeV$^2$.
	The contributions from the QE scattering alone
	(black dotted curves) are shown for comparison.}
\label{fig:arrington}
\end{figure}

The excellent agreement between the WBA model predictions and the
data holds over an even greater region of $Q^2$ than that shown
in Figs.~\ref{fig:lung1} and \ref{fig:lung2}.  Data from the
Jefferson Lab E89-008 experiment \cite{Arrington} spanning the
range $Q^2 \approx 1-7$~GeV$^2$ are also well reproduced by
the WBA model, as Fig.~\ref{fig:arrington} demonstrates.
The larger $Q^2$ coverage allows one to study the
relative importance of inelastic contributions at $x \sim 1$
compared with the QE.  While the cross sections are dominated
by QE scattering at $x \gtrsim 1$ for $Q^2 \lesssim 3$~GeV$^2$,
at higher $Q^2$ and lower $x$ [or larger $W^2 = (p+q)^2$]
the role of inelastic scattering from the nucleon becomes
increasingly more prominent.  To reproduce the full strength
of the inclusive cross section data in this region one must
therefore add the inelastic contribution to the QE.

The inelastic cross sections can be computed within the WBA
framework using the same smearing functions as those in
Eq.~(\ref{eq:fij}), convoluted with appropriate inelastic
free nucleon structure functions as in Eqs.~(\ref{eq:conv}).
A number of studies of inelastic deuteron structure functions
have previously been performed in the literature
\cite{KPW94, AKL04, KP06, Kahn09, AVQ}, and the smearing
functions have been used to extract information on the free
neutron structure function \cite{Malace10, Hen11, Rubin12},
and on parton distribution functions at large $x$ in global
QCD analyses \cite{ABKM09, CJ10, CJ11, CJ12, JMO13}.
Rather than repeat these analyses, for the purposes of the
present study it will be sufficient to simply employ the
inelastic contribution to the $F_1^d$ and $F_2^d$ structure
functions as parametrized in the phenomenological analysis
of Christy and Bosted \cite{CB10}.
As evident from Fig.~\ref{fig:arrington}, the inelastic
contributions become relevant at $x \lesssim 1$ for
$Q^2 \gtrsim 4$~GeV$^2$, although for $x \gtrsim 1$ or
$Q^2 \lesssim 2-3$~GeV$^2$ the cross sections are still
dominated by the QE component alone.

\begin{figure}
\includegraphics[width=17cm]{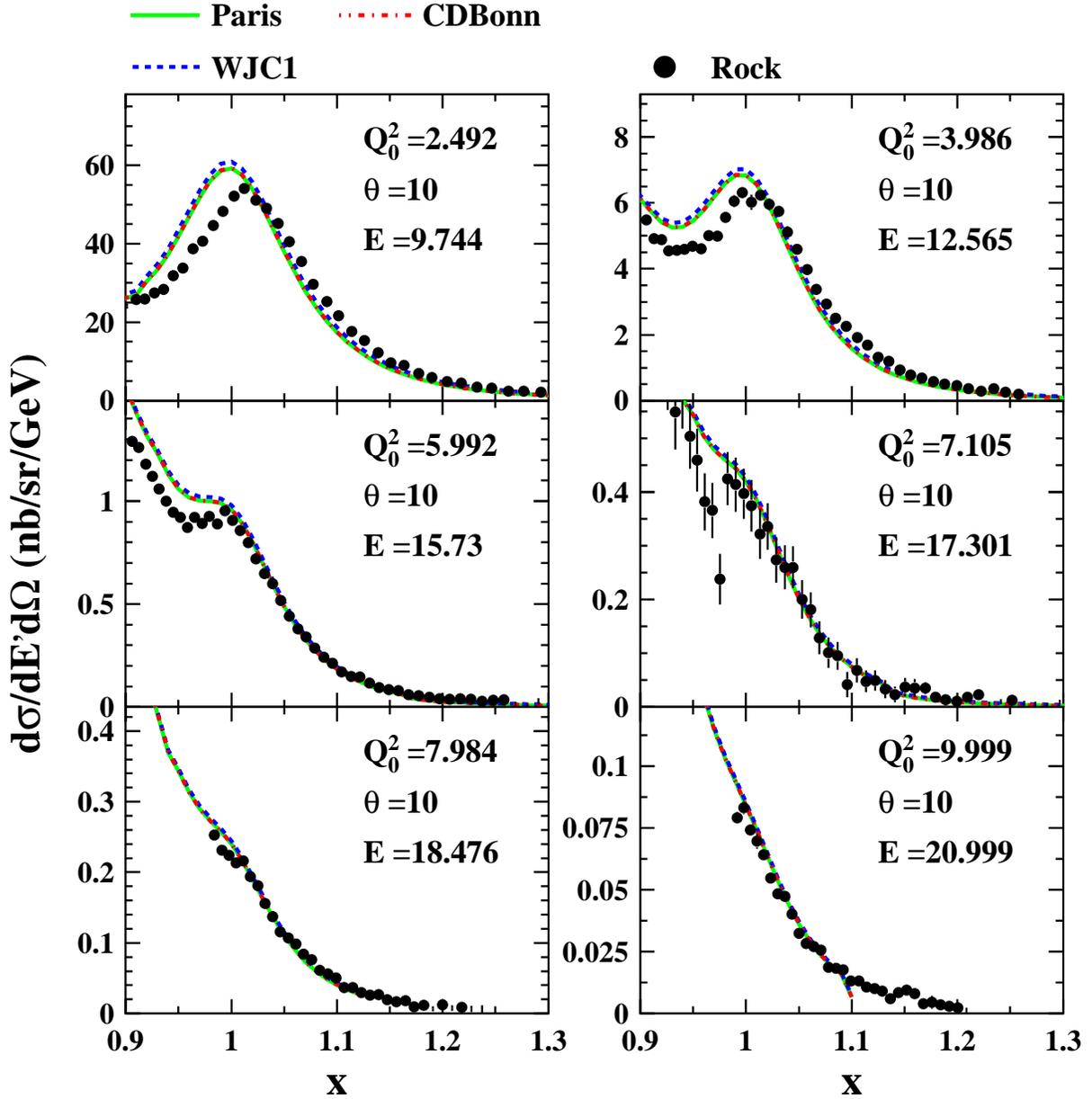}
\caption{As in Fig.~\ref{fig:arrington} but compared with the
	forward angle SLAC data from Rock {\it et~al.} \cite{Rock}
	at $\theta=10^\circ$, with $Q_0^2$ between $\approx 2.5$
	and 10~GeV$^2$.}
\label{fig:rock}
\end{figure}

Yet higher $Q^2$ values were reached in the earlier SLAC experiment
\cite{Rock} at small scattering angles ($\theta=10^\circ$), where
energies between $E \approx 10$ and 20~GeV allowed for $Q^2$ values
up to 10~GeV$^2$.
As Fig.~\ref{fig:rock} illustrates, once again the agreement is
generally good at $x > 1$, although curiously there appears a
mismatch in the position of the QE peak at $x \sim 1$, which is
most evident at the lower $Q^2$ values, $Q^2 \approx 2-4$~GeV$^2$.
This discrepancy appears difficult to reconcile with the otherwise
excellent agreement between the WBA model and data from other
experiments at SLAC \cite{Lung} and Jefferson Lab \cite{Arrington}
at similar kinematics, as evident in
Figs.~\ref{fig:lung1}--\ref{fig:arrington}
(see also Fig.~\ref{fig:fomin} below).
Note also that for the highest-$Q^2$ panel the theoretical curves
extend only to $x \approx 1.1$, corresponding to the maximum $Q^2$
values up to which the elastic form factors parametrizations are
given \cite{AMT07, Bosted94, Kelly04}.

\begin{figure}
\includegraphics[width=17cm]{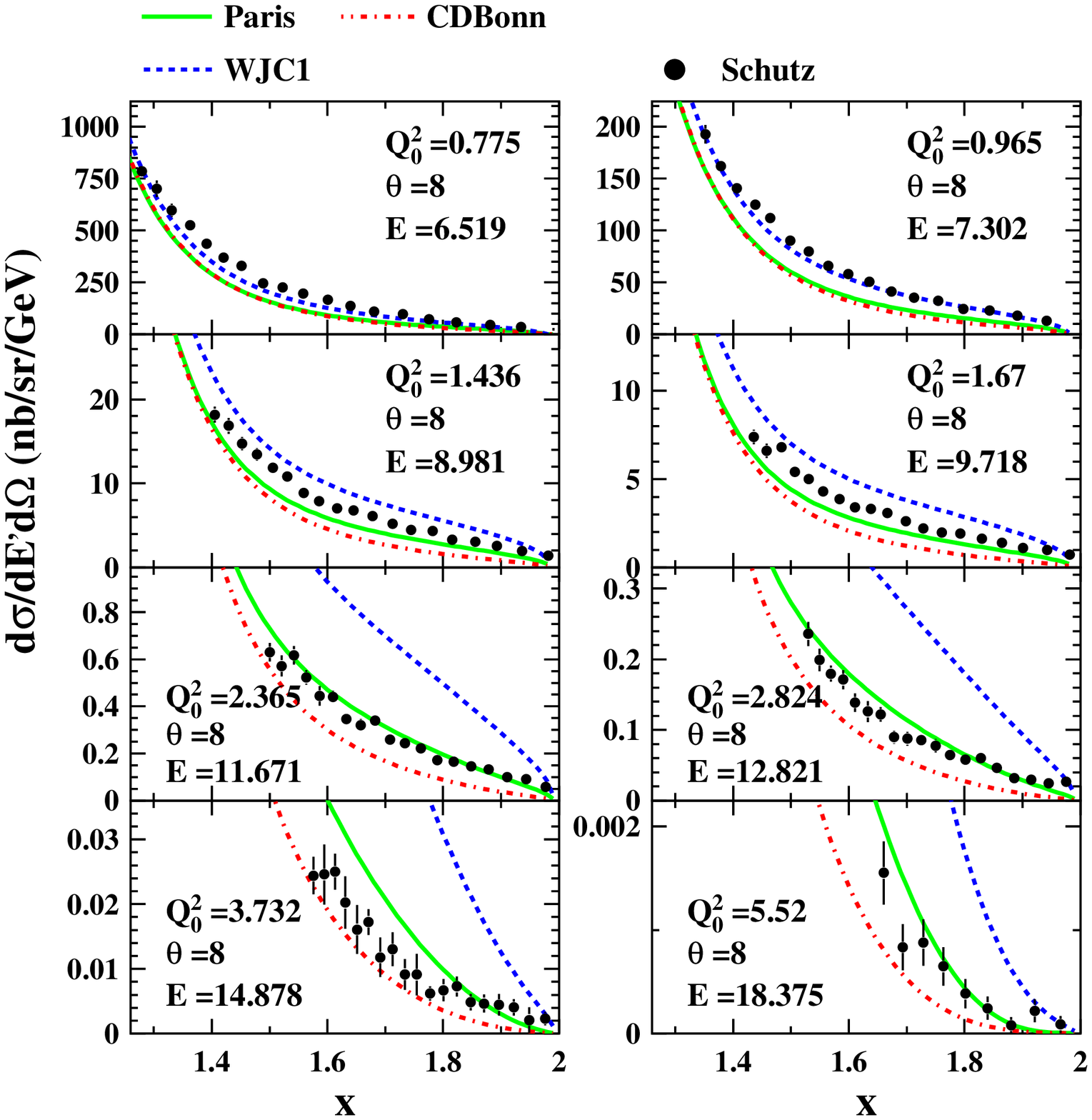}
\caption{As in Fig.~\ref{fig:arrington} but for the SLAC data from
	Schutz {\it et~al.} \cite{Schutz} at small scattering angles,
	for $Q_0^2$ ranging from $\approx 0.8$~GeV$^2$ to
	$\approx 5.5$~GeV$^2$.}
\label{fig:schutz1}
\end{figure}

At very high values of $x$ ($x \gg 1$), QE scattering from the
deuteron probes the tails of the smearing functions $f_{ij}(y)$
at $y \gg 1$.  As evident from Eq.~(\ref{eq:y}), large-$y$
kinematics is sensitive to large nucleon momenta $\bm{p}$, or
equivalently, to the short-range part of the $NN$ interaction
(see Fig.~\ref{fig:fy}).
Unlike the long distance component of the $NN$ potential,
which is well constrained by $pp$ and $pn$ scattering data,
the short-distance (or large-momentum) part of the deuteron
wave function has relatively large uncertainties.
This will translate into a larger spread in the theoretical
calculation of the deuteron structure functions when various
models for the wave function are used.
This is indeed observed in Fig.~\ref{fig:schutz1}, where data from
SLAC \cite{Schutz} at near-forward scattering angles are compared
with the QE cross sections computed using the Paris \cite{Paris},
WJC-1 \cite{WJC} and CD-Bonn \cite{CDBonn} wave functions.
As evident from the light-cone momentum distributions in
Fig.~\ref{fig:fy}, generally the CD-Bonn model gives rise to the
softest distribution, while the WJC-1 potential has the hardest
distribution, with the Paris wave function intermediate between these.
At the lower $Q^2$ values the data tend to prefer the harder
distributions, while softer wave functions are favored at
increasingly larger $Q^2$.

\begin{figure}
\includegraphics[width=17cm]{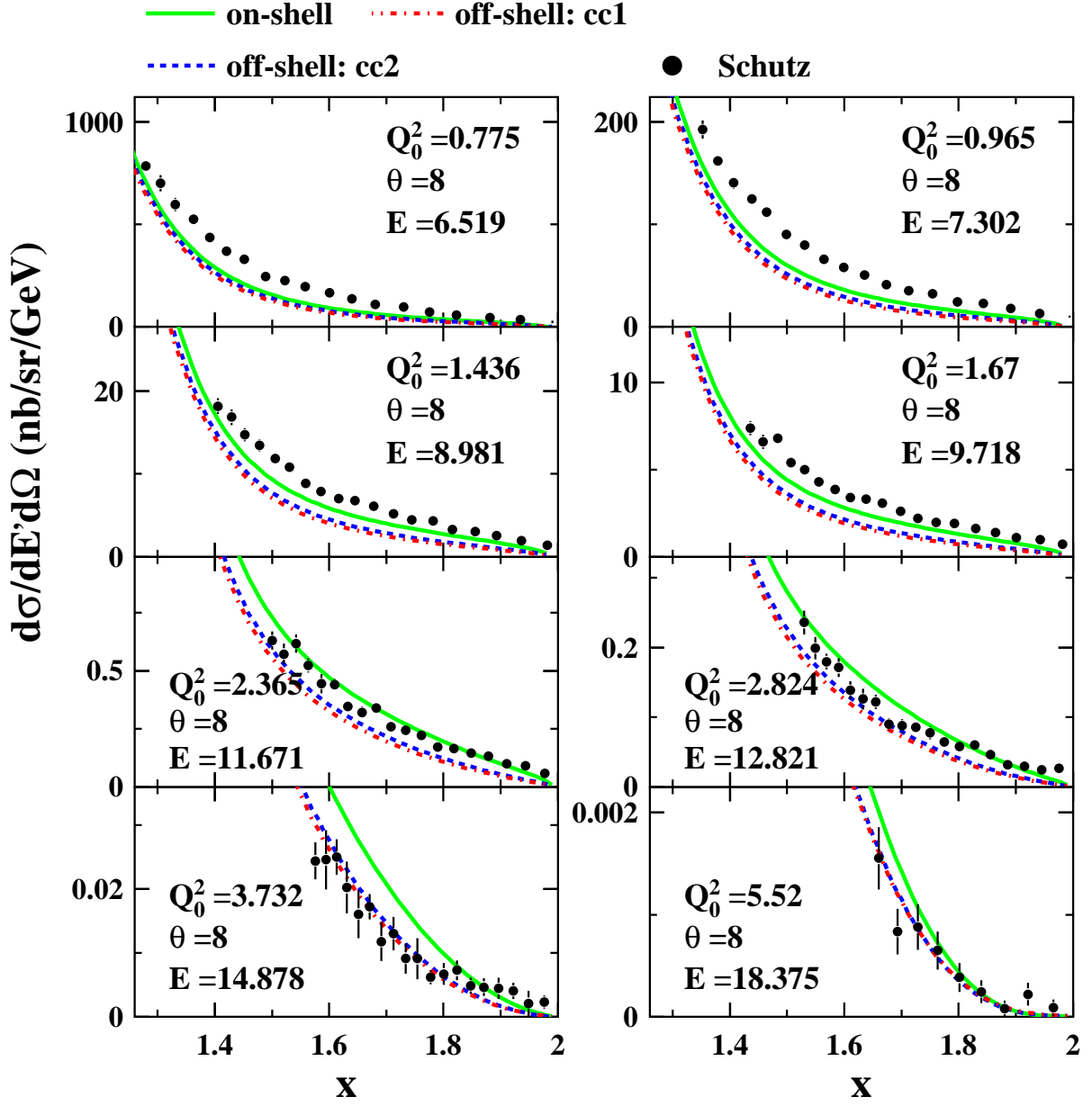}
\caption{Comparison of the WBA model predictions for the
	QE electron--deuteron at large $x$, using on-shell
	nucleon	form factors (green solid curves) and the two
	off-shell model extrapolations in Eqs.~(\ref{eq:off1})
	(off-shell ``cc2'', blue dashed curves) and (\ref{eq:off2})
	(off-shell ``cc1'', red dot-dashed curves).
	The Paris \cite{Paris} deuteron wave function is used in
	all cases, and the data are as in Fig.~\ref{fig:schutz1}.}
\label{fig:schutz2}
\end{figure}

In the same high-$x$ region where the uncertainties in the short-range
structure of the deuteron yield greater model dependence of the QE
cross sections, the effects of the possible off-shell dependence of
the nucleon elastic cross section are also expected to become more
important.
In Fig.~\ref{fig:schutz2} the WBA predictions for the cross sections
using on-shell nucleon form factors as in Eqs.~(\ref{eq:F12QE}) are
compared with calculations using the off-shell structure functions
from Eqs.~(\ref{eq:off1}) and (\ref{eq:off2}) in the generalized
convolution of Eqs.~(\ref{eq:F12QEful}).  For a meaningful comparison,
the Paris deuteron wave function is used for all cases.
The off-shell results with either the ``cc1'' or ``cc2'' models
generally soften the distributions relative to the on-shell
cross sections at high $x$, with the effects more pronounced
with increasing $Q^2$.
The off-shell corrections with the ``cc1'' model are slightly
larger in magnitude than those with the ``cc2'' model, although
the difference between these is significantly smaller than the
difference between the on-shell and off-shell results.

Compared with the high-$x$ Schutz {\it et~al.} data from SLAC
\cite{Schutz}, at the lower $Q^2$ values ($Q^2 \approx 1-2$~GeV$^2$)
the off-shell corrections with the Paris wave function make the
agreement slightly worse, confirming the findings in
Fig.~\ref{fig:schutz1} that these data prefer harder deuteron
wave functions.
In this region the WJC-1 wave function with minimal off-shell
corrections provides the best description of the data.
At higher $Q^2$ values ($Q^2 \approx 2-6$~GeV$^2$), using the
hardest, WJC-1 wave function would require significantly larger
off-shell corrections to reduce the excess of the calculated
cross section relative to the data.  The best agreement with data
here is obtained with the softer Paris wave function, together
with the off-shell nucleon form factors in Fig.~\ref{fig:schutz2}.
On the other hand, the softest wave function, with the CD-Bonn
potential, would underestimate the cross sections with the
addition of the off-shell nucleon corrections over all the
kinematics in Fig.~\ref{fig:schutz2}.

\begin{figure}
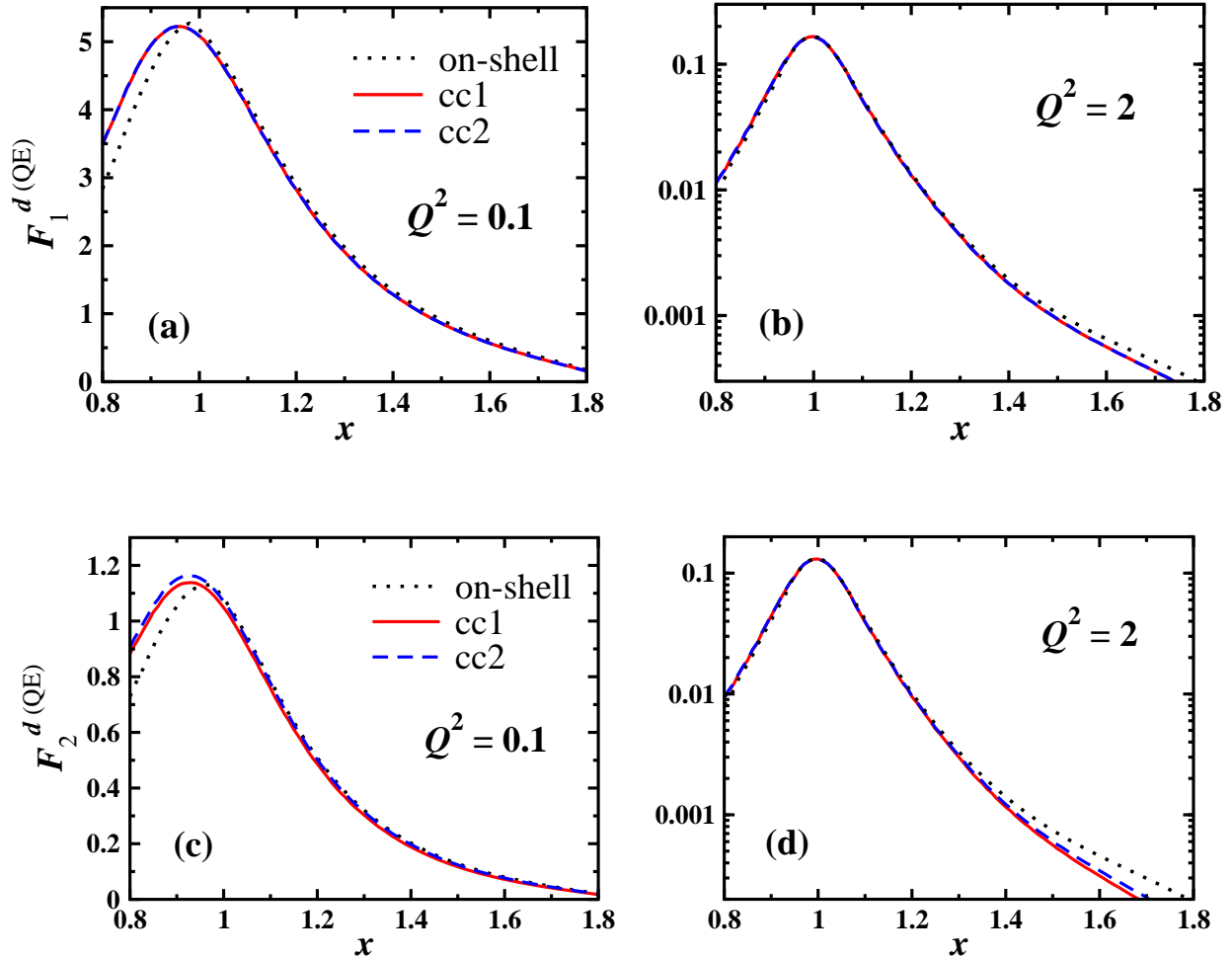

\includegraphics[width=8.0cm]{fig_F1_Q01.eps}\ \ \ \
\includegraphics[width=7.8cm]{fig_F1_Q2.eps}\vspace*{1.2cm}
\includegraphics[width=8.0cm]{fig_F2_Q01.eps}\ \ \ \
\includegraphics[width=7.5cm]{fig_F2_Q2.eps}
\caption{QE contributions to the deuteron $F_1^d$
	[{\bf (a)} and {\bf (b)}] and $F_2^d$
	[{\bf (c)} and {\bf (d)}] structure functions at
	$Q^2 = 0.1$~GeV$^2$ and $Q^2 = 2$~GeV$^2$.
	The on-shell approximation (black dotted curves)
	is compared with the off-shell calculation using the
	``cc1''	(red solid curves) and
	``cc2'' (blue dashed curves) prescriptions,
	with the Paris wave function used in all cases.}
\label{fig:F12}
\end{figure}

The behavior of the cross sections in Fig.~\ref{fig:schutz2}
can be understood from the effects of the off-shell corrections
on the $F_1$ and $F_2$ structure functions in Eqs.~(\ref{eq:off1})
and (\ref{eq:off2}).  In Fig.~\ref{fig:F12} the QE contributions
to the deuteron $F_1^d$ and $F_2^d$ structure functions with and
without off-shell corrections are shown at $Q^2 = 0.1$ and 2~GeV$^2$
for the ``cc1'' and ``cc2'' models.
Overall, the off-shell effects on the structure functions are
relatively small and weakly dependent on the choice of off-shell
prescription.
At low $Q^2$ ($Q^2 = 0.1$~GeV$^2$) the off-shell corrections are
noticeable only at $x \lesssim 1$, where they increase the magnitude
of the $F_1^d$ and $F_2^d$ structure functions by $\sim 10-20\%$.
At higher $Q^2$ values ($Q^2 = 2$~GeV$^2$), the off-shell effects
reduce the magnitude of the structure functions at high $x$
($x \gtrsim 1.4$), with a slightly larger correction appearing
for $F_2^d$ than for $F_1^d$, particularly for the ``cc1'' model.
This explains the suppression observed in the QE cross sections
at high $x$ and $Q^2$ in Fig.~\ref{fig:schutz2}, where the
forward angle data are dominated by the $F_2^d$ contribution
[see Eq.~(\ref{eq:sigma})].

\begin{figure}
\includegraphics[width=17cm]{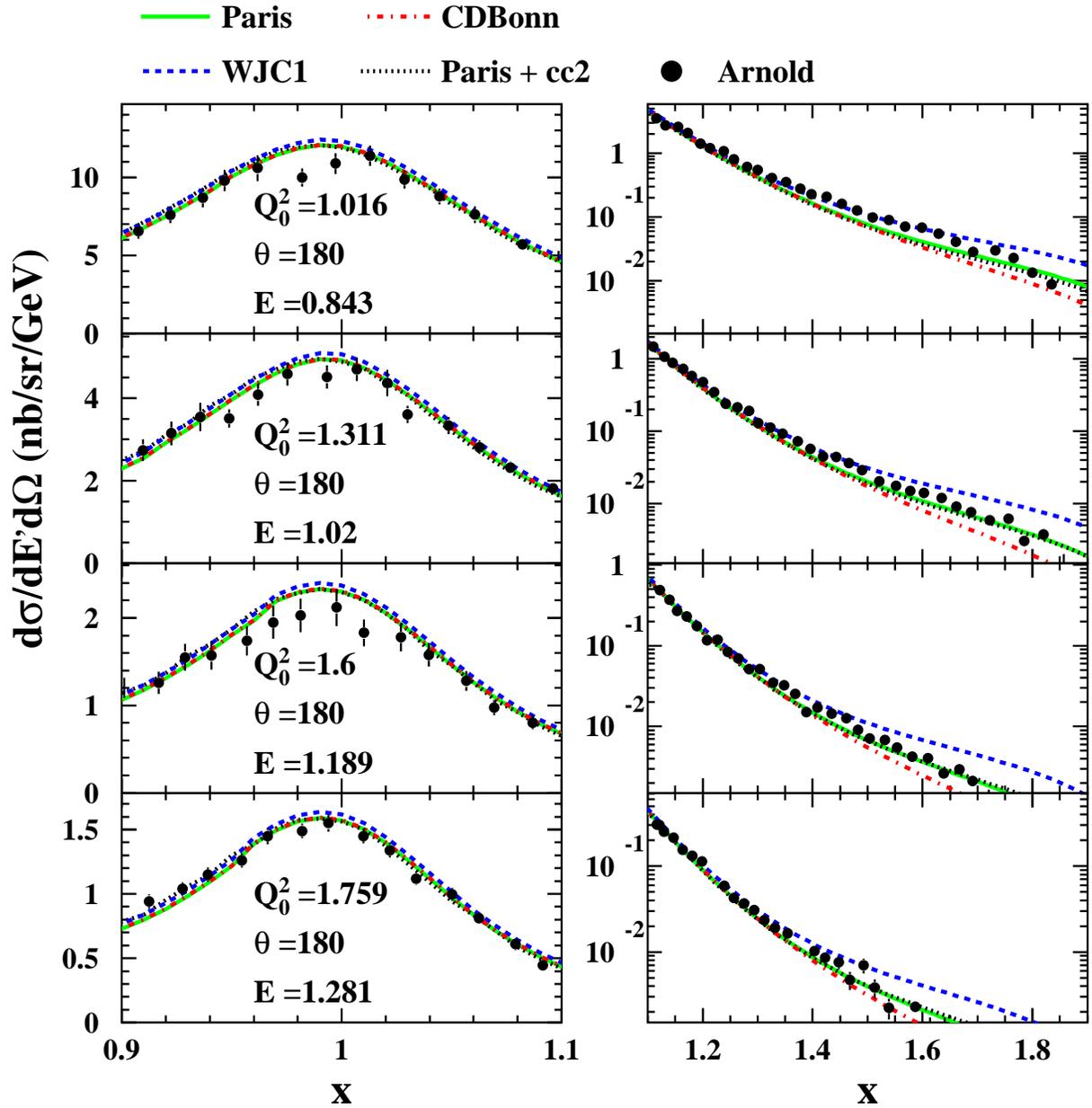}
\caption{Deuteron wave function and nucleon off-shell model
	dependence of the QE cross sections at backward angles
	for $Q_0^2$ between $\approx 1$ and 2~GeV$^2$.
	The on-shell results for the Paris (green solid curves),
	WJC-1 (blue dashed curves) and CD-Bonn (red dot-dashed curves)
	wave functions,	and the ``cc2'' off-shell model with the
	Paris wave function (black dot-dashed curves), are compared
	with the SLAC data from Arnold {\it et~al.} \cite{Arnold}.
	The left-hand panels illustrate the data in the vicinity
	of $x=1$ on a linear scale, while the right-hand panels
	show the tails of the cross sections at larger $x$ on a
	logarithmic scale.}
\label{fig:arnold}
\end{figure}

At extreme backward angles ($\theta=180^\circ$) the dominance of
magnetic scattering means that the cross section is given entirely
by the $F_1^d$ structure function.
Backward angle data from SLAC at $Q^2 \sim 1-2$~GeV$^2$ \cite{Arnold}
are compared in Fig.~\ref{fig:arnold} with WBA calculations over the
range $0.9 \lesssim x \lesssim 1.8$, including both deuteron wave
function and nucleon off-shell effects.
The overall agreement is very good, with the model dependence 
in the region of the QE peak, $0.9 \lesssim x \lesssim 1.1$,
essentially negligible.  (The results using the ``cc1'' off-shell
prescription are almost indistinguishable from those of the ``cc2''
model shown in Fig.~\ref{fig:arnold}.)
At larger $x$ values the wave function dependence becomes more
prominent, with the data at $x \lesssim 1.5$ better described
using the WJC-1 model, while the Paris wave function gives better
agreement at higher $x$.  The softer CD-Bonn wave function tends to
underestimate the data at the highest $x$, as observed for the
forward scattering angle data in Fig.~\ref{fig:schutz1}.
The off-shell corrections give a slight enhancement of the cross
section at $x \lesssim 1$, which is consistent with the behavior of
$F_1^d$ around the QE peak in Fig.~\ref{fig:F12}, but are otherwise
negligible at these kinematics.

\begin{figure}
\includegraphics[width=17cm]{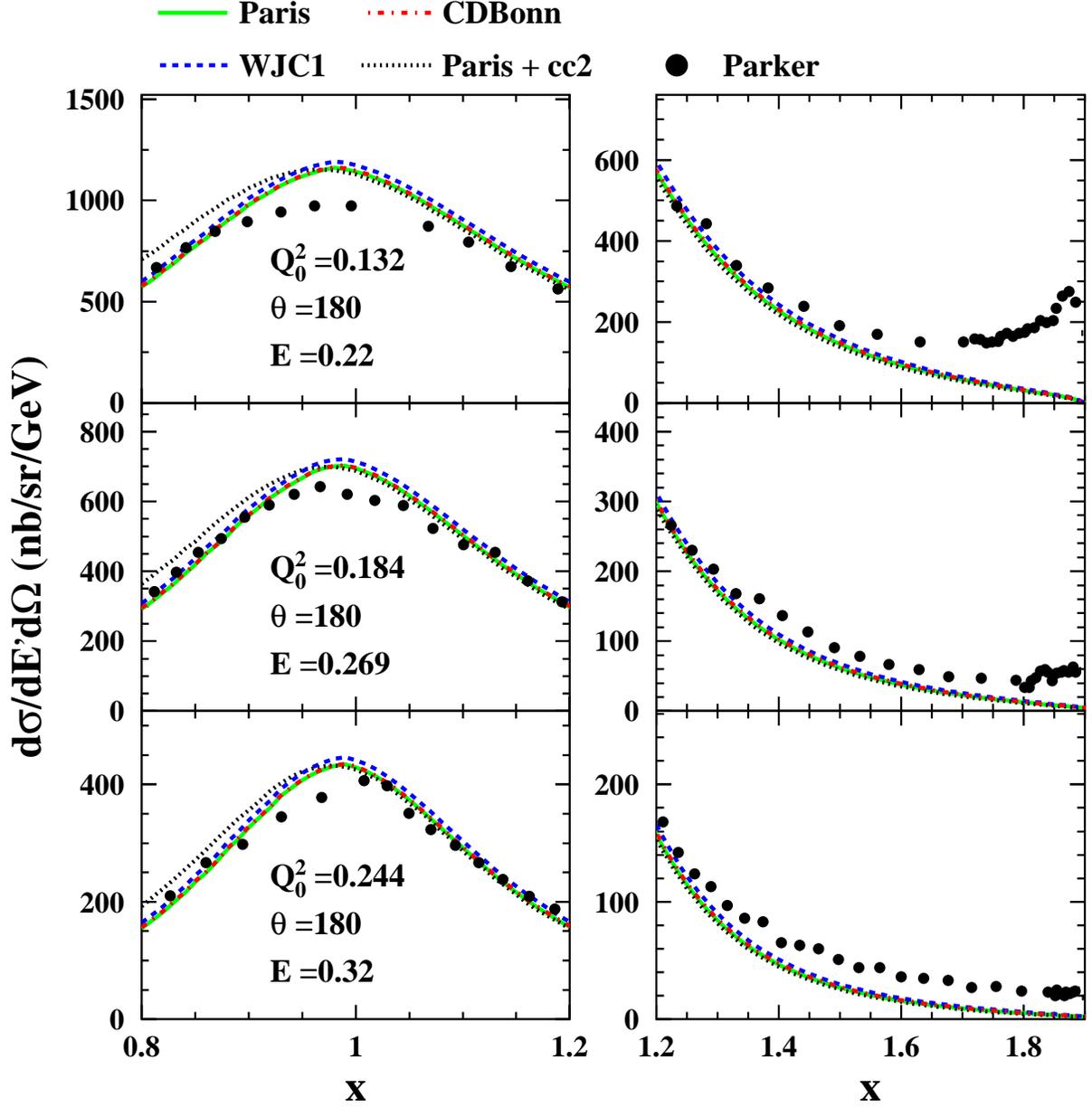}
\caption{As in Fig.~\ref{fig:arnold}, but for the lower-$Q^2$
	backward angle MIT-Bates data from Parker {\it et~al.}
	\cite{Parker}, for $Q_0^2 \sim 0.1-0.2$~GeV$^2$.}
\label{fig:parker}
\end{figure}

While a small, few percent enhancement of the backward angle QE
cross section at $x \lesssim 1$ due to off-shell effects is expected
from Fig.~\ref{fig:F12} at $Q^2 \approx 1-2$~GeV$^2$, since the
off-shell corrections in Eqs.~(\ref{eq:off1}) and (\ref{eq:off2})
scale with $(M^2-p^2)/Q^2$, the effects should be somewhat larger
at lower $Q^2$ values.
This is indeed observed in Fig.~\ref{fig:parker}, where low-energy
data from MIT-Bates \cite{Parker} at $Q^2 \sim 0.1-0.2$~GeV$^2$
indicate an $\approx 10\%-20\%$ enhancement at $x \approx 0.9$
compared with the on-shell cross section.  The cross sections with
the ``cc2'' off-shell model are displayed in Fig.~\ref{fig:parker}
(the results with the ``cc1'' model are again almost
indistinguishable), and the behavior follows directly from the
off-shell correction to $F_1^d$ at low $Q^2$ illustrated in
Fig.~\ref{fig:F12}.

At the low $Q^2$ values of the backward angle MIT-Bates data from
Parker {\it et~al.} in Fig.~\ref{fig:parker}, the dependence on the
deuteron wave function is very weak, even at large values of $x$.
All models appear to slightly overestimate the data in the
$x \sim 1$ region, possibly suggesting a role for meson exchange
currents at these kinematics.
Interactions between the virtual photon and a meson exchanged
between the two nucleons in the deuteron are known to affect the
$F_1$ structure function more so than the $F_2$ structure function
in QE electron--deuteron scattering \cite{MEC}.
The agreement between the calculations and data at $x \gtrsim 1$ is
very good, although at larger $x$ ($x \gtrsim 1.3$) the calculation
using the Paris wave function somewhat underestimates the data.
As observed for the forward angle data in Fig.~\ref{fig:schutz1},
here the harder momentum distribution associated with the WJC-1
deuteron wave function would produce better agreement.
As for the higher $Q^2$ backward angle data in Fig.~\ref{fig:arnold},
the off-shell corrections play a minor role in this region.
For even higher $x$, the data exhibit a significant rise as $x \to 2$,
especially at lower $Q^2$, which is likely due to the elastic
electron--deuteron scattering contribution, which drops rapidly
with increasing $Q^2$.

\begin{figure}
\includegraphics[width=17cm]{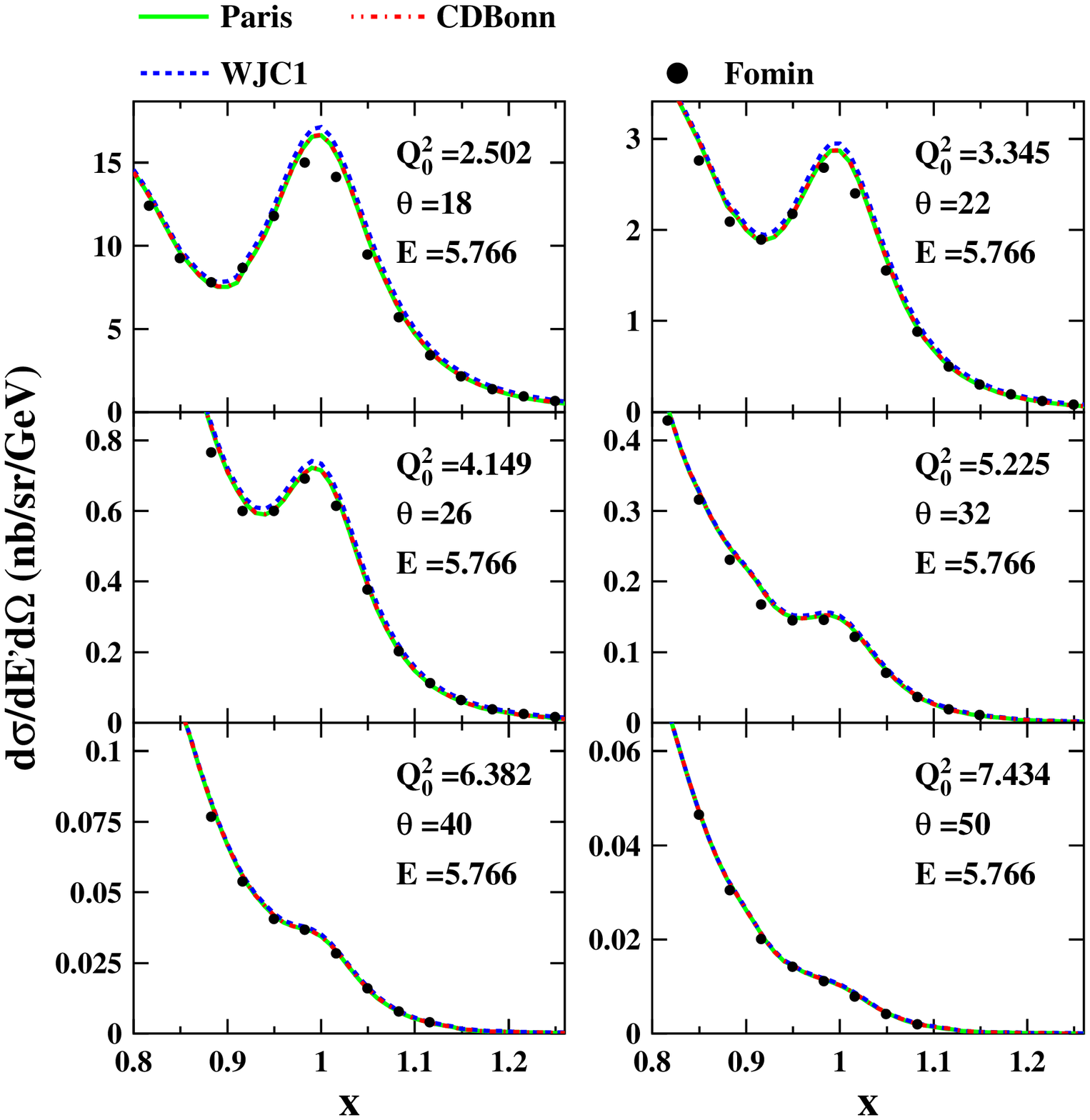}
\caption{Inclusive electron--deuteron QE scattering cross sections
	in the WBA model using the Paris \cite{Paris} (green solid
	curves), WJC-1 \cite{WJC} (blue dashed curves) and CD-Bonn
	(red dot-dashed curves) \cite{CDBonn} deuteron wave functions,
	compared with data from the E02-019 experiment in Hall~C
	at Jefferson Lab.  The incident energy is $E = 5.766$~GeV,
	with the scattering angles ranging from $\theta=18^\circ$
	to $50^\circ$, and $Q_0^2$ values from 2.5 to 7.4~GeV$^2$.}
\label{fig:fomin}
\end{figure}

Finally, the very latest and precise data on QE electron--deuteron
scattering from the Hall~C experiment E02-019 at Jefferson Lab
\cite{Fomin} are shown in Fig.~\ref{fig:fomin}, spanning a range
of $Q^2$ between $\approx 2$ and 8~GeV$^2$ and scattering angles
between $\theta \approx 18^\circ$ and $50^\circ$.
The agreement between the WBA model and the data is clearly excellent
over the complete $x$ range ($x \lesssim 1.25)$ covered, with very
mild dependence on the deuteron wave function.  The effects of nucleon
off-shell corrections are also negligible at these kinematics.
This close correspondence between the theory and experiment
provides further indication of the general success of the WBA
approach to describing inclusive electron--deuteron scattering.

\section{Conclusions}
\label{sec:conc}

We have performed a comprehensive analysis of QE electron--deuteron
scattering data within the framework of the weak binding approximation.
Using the same smearing functions for the bound nucleons in the
deuteron as those previously derived for deep-inelastic scattering
at finite $Q^2$, we have explored the limits of applicability of the
impulse approximation in the WBA.
Overall, we find excellent agreement between the model calculations
and the world's available data over a large range of kinematics,
covering $Q^2$ values between $\sim 0.1$ and 10~GeV$^2$, and $x$
values from below the QE peak to $x \approx 2$.
It is vital, however, that the correct kinematical $Q^2$ dependence
in the smearing function is taken into account in order to describe
the cross section data, in contrast to the high-$Q^2$ approximation
that can usually be assumed for deep-inelastic scattering.

The results are relatively independent of the details of the
deuteron wave function, except at very high values of $x$
($x \gtrsim 1.3$) and $Q^2 \gtrsim 1$~GeV$^2$, where there is
greater sensitivity to the high-momentum tails of the nucleon
momentum distributions in the deuteron.
For $Q^2 \sim 1$~GeV$^2$ the wave function based on the WJC-1
nucleon--nucleon potential \cite{WJC}, which has the hardest
momentum distribution, provides the best agreement with the QE
data, while for $Q^2 \gtrsim 2$~GeV$^2$ the Paris wave function
\cite{Paris} gives the best fit.
The CD-Bonn potential \cite{CDBonn}, with the softest momentum
distribution, tends to underestimate the data at the highest $x$
and $Q^2$ values.  This suggests that QE data at these kinematics
could be used to constrain the short-distance part of the $NN$
interaction, as reflected in the high-momentum behavior of the
smearing functions.

At high $x$ and low $Q^2$ corrections from nucleon off-shell effects
are also expected to play a role.  We considered two models for
extrapolating the nucleon electromagnetic current off-shell,
corresponding to the ``cc1'' and ``cc2'' prescriptions commonly
used in the literature \cite{DeForest83}.
Uncertainties in the off-shell corrections to structure functions
of nucleons in the deuteron is one of the main impediments to the
unambiguous extraction of the free neutron structure and the
determination of the $u$ and $d$ parton distribution functions at
large $x$ \cite{MST94, MSTPLB, Rubin12, ABKM09, CJ10, CJ11, CJ12}.
Studies of QE scattering can therefore provide additional
information on the off-shell corrections which could better
constrain the parton distribution function analyses.
In practice, we find relatively small off-shell corrections
for most kinematics, with the exception of very low $Q^2$
($Q^2 \sim 0.1-0.2$~GeV$^2$) at $x \lesssim 1$, where the
off-shell effects increase the on-shell cross sections, and
at very high $x$ ($x \gtrsim 1.4$) for $Q^2 \sim 1$~GeV$^2$,
where the cross sections are slightly reduced by the off-shell
effects.  The dependence on the off-shell prescription
(``cc1'' or ``cc2'') appears insignificant at the kinematics
where data currently exist.

In certain kinematic regions there are discrepancies between the
calculations and some of the data sets, such as at very low $Q^2$
values around the QE peak \cite{Parker}, where all calculations
overestimate slightly the data.  This may indicate a problem with
the data, or perhaps the need for additional corrections not
taken into account in this analysis.
Since the data in question \cite{Parker} are at extreme backward
angles, where $F_1^d$ dominates, this suggests that meson exchange
currents may play a role, as these are known to be more important
for the transverse response functions than for the longitudinal
\cite{MEC}.

In general, however, the WBA model provides a remarkably good
description of the QE data in all but the most extreme kinematics
($x \gg 1$ and $Q^2 \to 0$), which gives additional confidence
in the use of the finite-$Q^2$ smearing functions to compute
nuclear effects in other processes, such as inclusive
deep-inelastic scattering \cite{CJ10, CJ11, CJ12}.
In particular, the availability of QE data at both forward and
backward scattering angles allows the effects on the $F_1^d$
and $F_2^d$ structure function contributions to be studied
independently, and over a substantial range of $x$ and $Q^2$.
This poses a serious test of the model of the deuteron, and
provides clearer indications of the limits of applicability
of the WBA approach.

As far as the implications for future work, additional data at
high $x$ ($x \gtrsim 1.5$) and high $Q^2$, at forward and backward
angles, would be very helpful in constraining the model dependence
of the deuteron wave function, and possibly teasing out the
off-shell dependence of the nucleon structure functions.
On the theoretical front, inclusion of the QE deuteron data
in studies of $NN$ scattering could allow for a more reliable
determination of the large-momentum components of the deuteron
wave function.
For precision fits to the QE data, it will be necessary to explore
quantitatively in addition meson exchange currents, rescattering
(or final state interaction) effects, and the relativistic motion
of nucleons in the deuteron.  The present study should provide
an important baseline for these additional contributions.

\newpage
\begin{acknowledgments}

We thank D.~Day, S.~Kulagin, A.~Lung and J.~W.~Van~Orden for helpful
discussions and communications.
This work was supported by the U.S. DOE Contract No. DE-AC05-06OR23177,
under which Jefferson Science Associates, LLC operates Jefferson Lab.
N.D. acknowledges support from DOD's ASSURE program and NSF Award
No.~1062320 for an REU internship at ODU/Jefferson Lab.

\end{acknowledgments}


\end{document}